\documentclass[pre,preprint,showpacs,preprintnumbers,amsmath,amssymb]{revtex4}
\usepackage{graphics,epstopdf,epsfig,stmaryrd}

\newcommand{\ea}{\textit{et al.}}

\begin {document}
\title{The various manifestations of collisionless dissipation in wave propagation} \author{Didier B\'enisti}
\email{didier.benisti@cea.fr} \author {Olivier Morice}
\author{Laurent Gremillet}

\affiliation{ CEA, DAM, DIF F-91297 Arpajon, France.} \date{\today}
\begin{abstract}
The propagation of an electrostatic wave packet inside a collisionless and initially Maxwellian plasma is always dissipative because of the irreversible acceleration of the electrons by the wave. Then, in the linear regime, the wave packet is Landau damped, so that in the reference frame moving at the group velocity, the wave amplitude decays exponentially with time. In the nonlinear regime, once phase mixing has occurred and when the electron motion is nearly adiabatic, the damping rate is strongly reduced compared to the Landau one, so that the wave amplitude remains nearly constant along the characteristics. Yet, we show here that the electrons are still globally accelerated by the wave packet, and, in one dimension, this leads to a non local amplitude dependence of the group velocity. As a result, a freely propagating wave packet would shrink, and, therefore, so would its total energy. In more than one dimension, not only does the magnitude of the group velocity nonlinearly vary, but also its direction. In the weakly nonlinear regime, when the collisionless damping rate is still significant compared to its linear value, this leads to an effective defocussing effect which we quantify, and which we compare to the self-focussing induced by wave front bowing.  
\end{abstract}
   
\maketitle
\section{Introduction} 

Collisionless dissipation was first described by Landau in his famous paper Ref.~\cite{landau}~predicting the damping of an electrostatic wave, and is due to the global acceleration of the charged particles by the electrostatic potential, as was later discussed in Refs.~\cite{nicholson,elskens}. Landau damping is certainly one of the most important and basic features of wave propagation in a plasma, it has many applications (e.g.~current drive  in a tokamak, \cite{Fisch,Bernabei}), and has been extensively discussed in the scientific literature, which eventually led to its rigourous proof (see Ref.~\cite{mouhot}~and references therein). It is nevertheless not ubiquitous since, as shown by O'Neil in~Ref. \cite{oneil}, an electron plasma wave (EPW) with a large enough initial amplitude remains essentially undamped, because the rate of energy exchange between the electrons and the EPW  decreases on a timescale of the order of the bounce period, $T_B\equiv 2\pi \sqrt{m/eE_0k}$, $m$ being the electron mass, $-e$ its charge, $E_0$ the EPW amplitude and $k$ its wave number. Then, if $\omega_B \equiv 2\pi/T_B$ is much larger than the Landau damping rate, $\nu_L$, the EPW amplitude remains nearly constant (see Ref. \cite{oneil}). O'Neil's theory was the subject of some controversies (see Ref.~\cite{manfredi}~and references therein) but was checked experimentally in Ref.~\cite{danielson}~where it was shown that, if an EPW was excited with a small initial amplitude, $\omega_B < \nu_L$, it would damp at the rate predicted by Landau while, if $\omega_B \gg \nu_L$, after a few oscillations at a period close to $T_B$, the EPW amplitude would only slowly decrease on a timescale much larger than $\nu_L^{-1}$ due to some finite dissipation from the detection equipment. 

The nonlinear reduction of collisionless damping was recently found to have important implications in inertial confinement fusion, using the indirect drive approach \cite{lindl}, since it could lead to Raman reflectivities much larger than linear theory would predict on the grounds of Landau's prediction for the EPW damping rate. This was unambiguously shown experimentally in Ref.~\cite{montgomerry}, and these experimental results were qualitatively reproduced using kinetic simulations by several authors (for example in Refs.~\cite{strozzi,yin08}). The nonlinear reduction of the collisionless damping rate of an SRS-driven plasma wave was furthermore derived theoretically in Refs. \cite{benisti09,yampo}, two theories supported by Vlasov simulations and essentially equivalent over a finite range of wave amplitudes, as shown in Ref.~\cite{compare}. Moreover, in Refs.~\cite{PRL_10,brama}, by comparing results found in Vlasov simulations to those derived using a nonlinear envelope code following the theory of Ref.~\cite{benisti07}, levels of Raman reflectivity way beyond those predicted by linear theory were very clearly associated with the vanishing of the collisionless damping rate. Finally, anomalously large reflectivity levels measured at the National Ignition Facility~\cite{lindl},  and analyzed in Ref.~\cite{kirkwood}, seem to be due to nonlinear kinetic effects. 

Now, as a driven EPW grows in an initially Maxwellian plasma, it keeps on increasing the electron kinetic energy in an \textit{irreversible}~fashion, because it keeps on trapping more and more electrons. Moreover, the number of trapped electrons quickly increases with the EPW amplitude, and so does the amount of electrostatic energy which is dissipated. This seems to contradict the previous discussion on the nonlinear reduction of the collisionless damping rate which should vanish after a few bounce periods, and therefore more rapidly if the wave amplitude is large. One of the scopes of this paper is to resolve this apparent paradox by showing that an undamped wave may still be subjected to collisionless dissipation, which manifests itself in the nonlinear variations of the EPW group velocity, $v_g$.  In one dimension, $v_g$ decreases in that space region where the wave amplitude increases along its direction of propagation, and remains almost constant elsewhere. It is quite clear that this would automatically reduce the total energy of a freely propagating and undamped wave packet, simply because the total area located under the wave envelope would shrink, until a shock-like situation would occur (see Fig. \ref{f1}). Understanding this point allows us to discuss physically, in Section \ref{1D}, the results previously found in Ref.~\cite{vgroup}~on the nonlinear group velocity of an EPW. In particular, we clearly specify when, and how, collisionless dissipation enters to let us derive an envelope equation that does not follow from the variational approach developped by Whitham in Ref.~\cite{whitham}. Moreover, one important point of this paper is to show how the reasoning made in one dimension straightforwardly generalizes to allow for a multi-dimensional geometry. This lets us predict that collisionless dissipation leads to a non zero transverse component of the EPW group velocity that was quite unexpected, and which we very precisely quantify. In particular, we discuss its ability to offset, or to locally reinforce, the focussing effect due to the wavefront bowing induced by the EPW nonlinear frequency shift (see for example Ref.~\cite{yin08}) for a plasma wave resulting from stimulated Raman scattering. 

This paper is organized as follows. The next Section is devoted to one-dimensional wave propagation, starting with the situation when the wave amplitude is uniform. In this situation, we physically discuss the envelope equation derived previously in Refs.~\cite{benisti07,yampo}, and emphasize the very particular role played by the trapped electrons, which leads to an equation that would not follow from Whitham's theory. This envelope equation is then straightforwardly generalized to allow for a inhomogeneous wave amplitude, which lets us derive in an extremely simple fashion the results of Ref.~\cite{vgroup}~on the EPW group velocity. In particular, it is unambiguously shown that the nonlinear variations of $v_g$ allow for a collisionless dissipation mechanism of the electrostatic energy different from Landau damping. In Section~\ref{III}, the results obtained in one dimension are very easily generalized to multi-dimensional wave propagation, leading to new nonlinear focussing/defocussing effects which were, in some way, qualitatively discussed in Ref.~\cite{banks},  and which we quantify. This allows us to make comparisons with the focussing effect due to wave front bowing, and to discuss which of these effects prevails, depending on time and on space location. Section \ref{conc} concludes and summarizes our work. 

\section{One-dimensional wave propagation}
\label{1D}
\subsection{Uniform wave amplitude}
In this Subsection, we consider a nearly monochromatic electron plasma wave, with frequency $\omega$ and wave number $k$, whose amplitude varies slowly in time while remaining space independent. We moreover restrict here to the situation when the time it takes for an electron to complete one ``frozen'' orbit (i.e., corresponding to a fixed wave amplitude) is much smaller than the typical timescale of variation of the EPW amplitude. For a harmonic wave, this condition is achieved when $\omega_B \gg \gamma$, where $\gamma$ is the EPW growth rate. In this case, one may assume that the electron orbits are infinitely close to the ``frozen'' ones, and that these orbits are completed in an infinitely small time (which would correspond to making use of the adiabatic approximation). In other words, the electron motion is ``enslaved'' to the variations of the wave amplitude. Then, in this situation, it is clear that any observable related to the electrons (energy, momentum, charge density,\dots) would vary at a rate nearly proportional the the wave growth rate, $\gamma$. As for the ions, for the sake of simplicity, we henceforth assume that they are immobile.

\begin{figure}
  \centerline{\includegraphics[width=17cm]{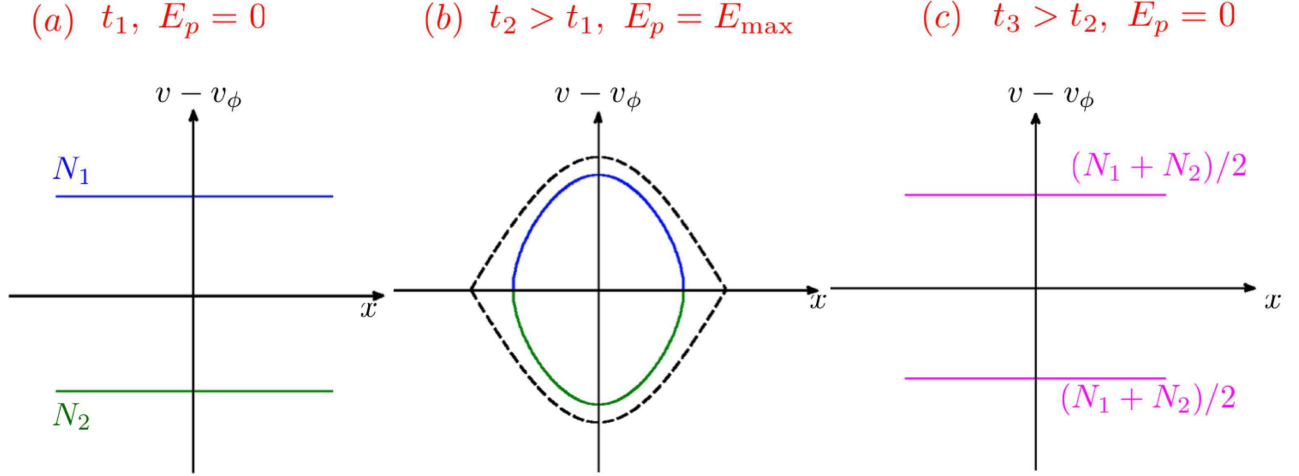}}
   \caption{\label{f0} (Color online) Space portrait of electrons with initial distribution function $f_i=N_1\delta(v-v_\phi-\Delta v)+N_2\delta(v-v_\phi+\Delta v)$ acted upon by an electrostatic wave. (a) The wave amplitude is zero. (b) The wave has grown slowly enough to induce adiabatic motion and has trapped the electrons which, due to action conservation, all lie on the same orbit. (c) In the limit when the period of the trapped orbit is infinitely small compared to the typical timescale of variation on the EPW amplitude, when this wave decays, half electrons are detrapped with $v>v_\phi$ and half with $v<v_\phi$. Then, again due to action conservation, when the EPW amplitude has decreased back to zero the distribution function is $f_0=[(N_1+N_2)/2][\delta(v-v_\phi-\Delta v)+\delta(v-v_\phi+\Delta v)]$.}
\end{figure}

Let us now consider an EPW whose amplitude, $E_p$, grows from  $E_{\min}$ to $E_{\max}$, and then decreases back to $E_{\min}$. If the electron motion has kept on being adiabatic during these variations of $E_p$, it is easily shown that the total energy of the electrons which have never been trapped is only a function of the EPW amplitude. Hence, when the wave grows, the gain in energy of the untrapped electrons is reversible and given back to the wave as it decays. Now, whatever $\Delta v$, adiabatic electrons with initial velocities $v_\phi \pm \Delta v$, where $v_\phi \equiv \omega/k$ is the EPW phase velocity, are all trapped at the same time, and their trapping leads to a jump in the kinetic energy by a quantity proportional to 
\begin{equation}
\label{4}
\Delta \mathcal{E}=m[f_0(v_\phi-\Delta v)-f_0(v_\phi+\Delta v)]v_\phi \Delta v,
\end{equation}
where $f_0$ is the electron distribution function in the limit of a vanishing wave amplitude. Note that $\Delta \mathcal{E}$ is non zero only because $f_0(v_\phi-\Delta v)\neq f_0(v_\phi+\Delta v)$ and, actually, $\Delta \mathcal{E}$ is positive when $f_0$ is a decreasing function of velocity, as is the case for a Maxwellian. Now, as shown theoretically in Ref.~\cite{benisti07}, and numerically in Refs.~\cite{brunner,vlasovia}, while $E_p$ is decreasing, the electrons are detrapped nearly symmetrically with respect to the phase velocity (see Fig.~\ref{f0} for a sketch of symmetric detrapping). Consequently, detrapping would lead to a change in the electron kinetic energy by $\Delta \mathcal{E}'=-\Delta \mathcal{E}$ but, now, with $f_0(v_\phi-\Delta v)= f_0(v_\phi+\Delta v)$. Hence, $\Delta \mathcal{E}'=0$, the energy gained by the electrons through trapping is not given back to the wave when they are detrapped. We therefore conclude that, \textit{when the electron motion is nearly adiabatic, only trapping may lead to an irreversible increase of the kinetic energy, and therefore to the collisionless dissipation of the electrostatic energy}. This point is actually not new and was already discussed in Refs.~\cite{fahlen,banks}. However, the scope of this paper is to describe in detail the implications of collisionless dissipation in the limit when the electron motion is nearly adiabatic, and to make clear how it may differ from Landau damping, which is one distinction we could not find in any of the papers we know of. Our discussion will be based on the envelope equation for the EPW amplitude, $E_p$.

Such an equation was derived by Yampolsky and Fisch in Ref.~\cite{yampo}~for a \textit{growing}~driven wave, and these authors found
\begin{equation}
\label{1}
\partial_\omega \chi_r(\partial_t +\nu)E_p=E_d,
\end{equation}
where $\chi_r$ is the real part of the electron susceptibility, $E_d$ accounts for the effect of the drive, while $\nu$ is directly related to the rate of energy gained by the electrons. As discussed before, for adiabatic electrons, the rate of energy variation is proportional to the EPW growth rate, and therefore so should be $\nu$ in the limit when $\omega_B \gg \gamma$. This is exactly the result found by Yampolsky and Fisch in Ref.~\cite{yampo}~so that, when  $\omega_B \gg \gamma$, one may write $\nu \partial_\omega \chi_r E_p \equiv \partial_\omega \chi'_r \partial_t E_p$, where  $\chi'_r$ is a dimensionless function of the EPW amplitude, and Eq.~(\ref{1}) becomes
\begin{equation}
\label{2}
(\partial_\omega \chi_r+\partial_\omega \chi'_r)\partial_t E_p=E_d.
\end{equation}
Unlike in Ref.~\cite{yampo}, in this paper we make the choice to no longer term $\nu$ the ``nonlinear Landau damping rate of the EPW'' when Eq.~(\ref{1}) is essentially equivalent to Eq.~(\ref{2}). Indeed, when this condition is met and when the EPW is freely propagating ($E_d=0$), its amplitude remains constant, the wave is undamped. Yet, if the wave is driven and keeps on trapping new electrons, it increases the electron energy  in an \textit{irreversible} fashion, and is therefore subjected to collisionless dissipation, which is accounted for by the term $\partial_\omega \chi'_r\partial_t E_p$ in Eq.~(\ref{2}). This is a first illustration of the distinction we make between Landau damping and collisionless dissipation. 

It is very interesting to note that the same kind of envelope equation as Eq.~(\ref{2}) was obtained by B\'enisti~\ea~in Refs.~\cite{benisti07,brama}~without ever calculating the rate of energy gained by the electrons. Indeed, starting from Gauss law, B\'enisti~\ea~straightforwardly find $\chi_i E_p=E_d$, where $\chi_i$ is the imaginary part of the electron susceptibility. Moreover, using an argument resting on the orbit symmetry, these authors show that, in the limit $\omega_B \gg \gamma$, the contribution to $\chi_i$ of the trapped electrons is negligible, so that $\chi_i \approx \chi_i^{\text{untr}}$, where $\chi_i^{\text{untr}}$ is the value of $\chi_i$ obtained by accounting only for the untrapped electrons. For a slowly varying wave amplitude, it is easily shown that $\chi_i^{\text{untr}}$ is nearly proportional to the wave growth rate, which leads to the following envelope equation,
\begin{equation}
\label{3}
\partial_\omega \chi_r^{\text{untr}}\partial_t E_p=E_d,
\end{equation}
where $\chi_r^{\text{untr}}$ is the sole contribution of the untrapped electrons to the real part of the susceptibility. Since, as shown in Ref.~\cite{compare}, the results obtained by B\'enisti~\ea~and Yampolsky and Fisch match over a range of wave amplitudes where the condition $\omega_B \ggÊ\gamma$ holds, Eq.~(\ref{3})~may be identified with Eqs.~(\ref{1}) and (\ref{2}). This shows that, when $\omega_B \ggÊ\gamma$, $\nu/\gamma \approx -\partial_\omega \chi_r^{\text{tr}}/\partial_\omega \chi_r$, where $\chi_r^{\text{tr}}$ is that part of $\chi_r$ only due to the trapped electrons. This formally relates collisionless dissipation to trapping, as expected from our previous discussion.

The physics content of Eq.~(\ref{1}) seems extremely clear, the rate of variation of the wave amplitude results from the balance between the energy gained from the drive and that given to the electrons. When this equation may be cast in the form of Eq.~(\ref{2}) or Eq.~(\ref{3}) then, as shown in Refs.~\cite{benisti07,vgroup}, $\partial_\omega \chi_r^{\text{untr}}$ is larger than $\partial_\omega \chi_r$, all the more as the (linear) Landau damping rate of the EPW is large. This just reflects the fact that a driven EPW would grow slower if a large fraction of the energy it gains from the drive is given back to the electrons through trapping. 

Note, though, that the theory by Yampolsky and Fisch~\cite{yampo}, based on energy conservation, only holds when the EPW keeps growing. By contrast, the argument on the symmetry of the trapped orbits given by B\'enisti \ea~in Ref.~\cite{benisti07}~is clearly valid whether the wave grows or decays. Hence, provided that the electron motion is nearly adiabatic, $\chi_i \approx \chi_i^{\text{untr}}Ê\approx \gamma \partial_\omega  \chi_r^{\text{untr}}$, so that Eq.~(\ref{3}) holds independently of the sign of $\partial_t E_p$. Nevertheless, although this equation is always formally valid, for a given $E_p<E_{\max}$, $\partial_\omega \chi_r^{\text{untr}}$ does not assume the same values when $E_p$ has kept on growing as when $E_p$ has previously decayed from $E_{\max}$. Indeed, as shown in Ref.~\cite{benisti07}, due to symmetric detrapping, the contribution to $\partial_\omega \chi_r^{\text{untr}}$ of the \textit{detrapped} electrons is negligible. Hence, Eq.~(\ref{3}) would be better written
\begin{equation}
\label{5}
\partial_\omega \chi_r^{\text{eff}} \partial_t E_p=E_d,
\end{equation}
where $\chi_r^{\text{eff}}$ is calculated by accounting only for the electrons which have \textit{never} been trapped. The envelope equation for the plasma wave is therefore non-local in the EPW amplitude since the value of $\partial_\omega \chi_r^{\text{eff}}$ at time $t$ does not only depend on $E_p(t)$ but also on $\max_{t'<t}[E_p(t')]$. As shall be seen in the next Subsection, this non-locality  entails the reduction of the total energy of a freely propagating wave packet. Moreover, when the wave is essentially undamped (i.e., when its amplitude remains constant along the characteristics), this is the main cause for the dissipation of the electrostatic energy. 

From the previous discussion we conclude that, if one were to write a Lagrangian for the wave-particle interaction, this Lagrangian would be non-local in $E_p$, so that it would not be quite easy to derive the envelope equation for the EPW from a variational approach, and this program will not be pursued in this paper. However, it is very interesting to discuss physically the correction that needs to be made to the variational approach developed by Whitham in Ref.~\cite{whitham}~in order to recover Eq.~(\ref{5}). In Whitham's theory, the Lagrangian is $\mathcal{L}(E_p)=\int_0^{E_p} [1+\chi_r(E')]E'dE'$ %\cite{note0}~
(notice that this is the Lagrangian one would derive within Whitham's formalism which does not account for dissipation. This is not the Lagrangian that should be used for the nonlinear propagation of an EPW we are addressing here). Then, for a freely propagating wave whose amplitude is uniform, since Whitam's theory cannot account for dissipation, it would trivially predict $\partial_\omega \chi_r \partial_t E_p=0$. For a driven wave, this equation would generalize into $\partial_\omega \chi_r \partial_t E_p=E_d$. Based on this, we now rewrite Eq.~(\ref{5}) as
\begin{equation}
\label{6}
\partial_\omega \chi_r \partial_t E_p+\partial_\omega [\chi_r^{\text{eff}}- \chi_r] \partial_t E_p=E_d. 
\end{equation}
The second term in the left-hand side of the previous equation clearly is  the correction that must be made to Whitham's theory in order to accurately describe the nonlinear time evolution of the EPW amplitude, and its physics content is transparent. Indeed, $(\chi_r^{\text{eff}}- \chi_r)$ is the contribution to the real part of the susceptibility of those electrons which have been trapped at least once, while $\partial_t E_p$ is proportional to the trapping rate. Hence, the correcting term is directly related to trapping, because in the limit $\omega_B \gg \gamma$ only trapping entails collisionless dissipation.  Understanding this will let us generalize Eq.~(\ref{6}), in a very simple and powerful fashion, in order to allow for a multi-dimensional variation of the wave amplitude, as we shall show it in the remainder of this paper.

Before this, we would like to stress that collisionless dissipation, and not trapping by itself, is at the origin of the correcting term in Eq.~(\ref{6}). Indeed, consider again an EPW whose amplitude grows from  $E_{\min}$ to $E_{\max}$, and then decreases back to $E_{\min}$. If the initial distribution function is symmetric with respect to  $v_\phi$ on a given velocity interval, $I_v$, then, from Eq.~(\ref{4}), the trapping of electrons with initial velocities in $I_v$ does not change the total kinetic energy. Moreover, those electrons, either trapped or not, do not contribute to $\partial_\omega \chi_r$. Hence, as long as only those electrons have been trapped, there has been no irreversible increase in the electron kinetic energy, and the correcting term in Eq.~(\ref{6}) is strictly zero although trapping did occur. 

\subsection{Non-uniform wave amplitude}

In the previous Subsection, we concluded that Whitham's result had to to be corrected to account for collisionless dissipation which, when the electron motion is nearly adiabatic, is only due to trapping. Using this point, it is extremely easy to generalize Eq.~(\ref{6}) to a wave whose amplitude only varies along its direction of propagation, $x$. Indeed, to do so, we only need to add to the left-hand side of Eq.~(\ref{6}) the term $-\partial_k \chi_r \partial_xE_p$ coming from Whitham's theory~\cite{note}, and to express the fact that the correction to Whitham's result is proportional to the rate of electron trapping, and therefore proportional to the EPW growth rate \textit{calculated in the frame moving at the phase velocity with respect to the laboratory frame}. Hence, this term is $\partial_\omega [\chi_r^{\text{eff}}- \chi_r] [\partial_t E_p+v_\phiÊ\partial_x E_p]$ so that, once Landau damping has vanished, the nonlinear envelope equation of the EPW is
\begin{equation}
\label{7}
\frac{\partial \chi_r}{\partial \omega} \frac{ \partial E_p}{\partial t}-\frac{\partial \chi_r}{\partial k} \frac{\partial E_p}{\partial x} +\frac{\partial [\chi_r^{\text{eff}}- \chi_r]}{\partial \omega} \left[\frac{\partial E_p}{\partial t}+v_\phiÊ\frac{\partial E_p}{\partial x} \right]=E_d.
\end{equation}
This is exactly the result found in Ref.~\cite{vgroup}, which was derived in the Appendix of that paper directly from Gauss law and not by using a variational approach. In Ref.~\cite{vgroup}, we already discussed physically that our result differed from the one derived by Whitham because of collisionless dissipation. Here, we will push the reasoning further and we will show in a few lines that the nonlinear variation of the EPW group velocity, as derived in Ref.~\cite{vgroup}, does entail the collisionless dissipation of the electrostatic energy. Here, we call
\begin{equation}
\label{8}
v_g \equiv -\frac{\partial_k \chi_r}{\partial_\omega \chi_r^{\text{eff}}}+\frac{\partial_\omega [\chi_r^{\text{eff}}- \chi_r]}{\partial_\omega \chi_r^{\text{eff}}} v_\phi
\end{equation}
the EPW group velocity because, if the wave was freely propagating, its amplitude would remain constant along the lines $x=v_gt$ \cite{note}. From the results of the previous Subsection, we clearly understand that $v_g$ differs from $-\partial_k \chi_r/\partial_\omega \chi_r$ because of the collisionless dissipation entailed by trapping.  

\begin{figure}
 \centerline{\includegraphics[width=13cm]{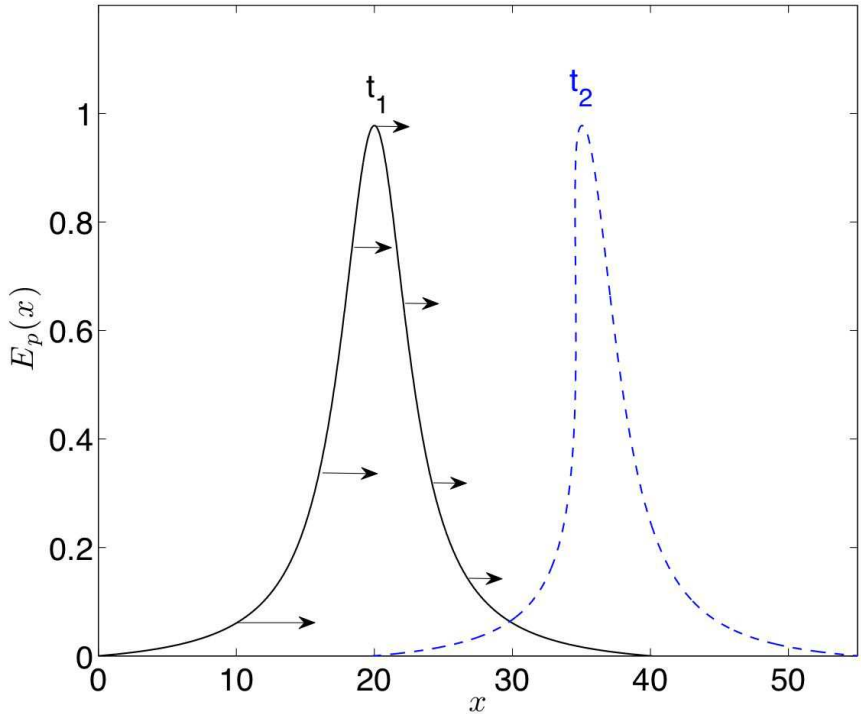}}
 \caption{\label{f1} (Color online) Space profile of the plasma wave (in arbitrary units), at time $t_1$ (black solid line) and at time $t_2>t_1$ (blue dashed line), when the group velocity of the wave packet (whose amplitude is indicated by the arrows) decreases with the EPW  amplitude at the rear side, and remains fixed at its minimum nonlinear value at the front side. One clearly sees that the wave packet at time $t_2$ is narrower than at time $t_1$.}
\end{figure}
Now, in order to discuss the implications of the nonlinear variations of $v_g$ in the decrease of the total electrostatic energy, we first consider a freely propagating wave packet whose evolution is given by Eq.~(\ref{7}) with $E_d=0$ (we therefore consider here a strongly nonlinear regime, where we neglect the effect of damping which will be discussed below, and which is assumed to be localized in a space region where the EPW amplitude is very small compared to its maximum value). Since the wave packet is essentially undamped, its maximum amplitude remains nearly constant. Moreover, as shown in Ref.~\cite{vgroup}, $v_g=v_\phi-2/(k\partial_\omega \chi_r^{\text{eff}})$, so that $v_g<v_\phi$. Hence, along the rear side of the wave packet, i.e. before its maximum, $E_p(x+v_\phi t)$ keeps on increasing so that, as shown in  Ref.~\cite{vgroup}, $v_g$ decreases (for the sake of simplicity, we assume here that the wave packet has no local maxima). In front side of the wave packet $E_p(x+v_\phi t)$ keeps on decreasing, so that $\max_{t'<t}\{E_p[x-v_\phi (t-t'),t']\}$ remains constant and so does the number of electrons which have been trapped by the EPW. Since $\partial_\omega \chi_r^{\text{eff}}$ essentially depends on these parameters, the group velocity remains nearly fixed at its minimum nonlinear value along the whole front side of the wave packet. Hence, the rear side moves faster than the front side so that the wave packet shrinks while its maximum amplitude remains constant, as illustrated in Fig.~\ref{1}, which automatically reduces the total electrostatic energy, $W \equiv \int_{-\infty}^{+\infty} E_p^2dx$. This goes on, at most, until a shock would occur in the wave profile. At this point, the EPW amplitude, as seen by the electrons, would no longer vary slowly so that the adiabatic approximation would break down and our analysis would no longer hold. What happens then is outside the scope of this paper and will not be addressed here.

Our prediction on the shrinking of the wave packet is consistent with the numerical results of Ref.~\cite{fahlen}. However, that paper only discusses the inhomogeneity of the Landau damping rate, larger at the rear side where the amplitude should therefore decrease more rapidly (or increase less rapidly for a driven wave) than at the front side, which leads to an effective shrinking of the wave packet. Here we show that another effect, the inhomogeneity of the EPW group velocity, which is one manifestation of collisionless dissipation but which occurs only after Landau damping has nearly vanished, may also reduce the space extent of the electrostatic wave packet.

In order to clarify the role of each of these effects, let us consider a freely propagating EPW whose envelope equation is (see Ref.\cite{brama})
\begin{equation}
\label{9}
\partial_t E_p +v_g \partial_x E_p +\nu E_p =0,
\end{equation}
where $\nu$ is what we call the nonlinear Landau damping rate of the EPW. Using an integration by parts, one easily finds
\begin{equation}
\label{10}
\partial_t \int_{-\infty}^{+\infty} E_p^2dx=-2\int_{-\infty}^{+\infty} \nu E_p^2 dx+\int_{-\infty}^{+\infty}\partial_x v_g E_p^2 dx.
\end{equation}
As shown in Ref.~\cite{benisti09}, $\nu$ is a decreasing function of $Y\equiv \int_{-\infty}^{t} \omega_B(x-v_\phi t',t') dt'$. Since $v_\phi>v_g$, $\nu$ is indeed larger at the rear side of the wave packet than at the front side as argued in Ref.~\cite{fahlen}. Actually, in Refs.~\cite{benisti09,vgroup} we showed that $\nu(x,t)$ and $v_g(x,t)$ remained very close to their linear values where $Y(x,t)\alt 6$, making, there, Landau damping the main cause of collisionless dissipation. In the region where $Y(x,t) \approx 6$, $\nu(x,t)$ quickly decreases with $x$, while $v_g$ quickly increases towards $v_\phi$. Hence, in the small space region where $Y \approx 6$ dissipation is, again, only due to Landau damping while the effect of $\partial_x v_g$ is anti-dissipative. Actually, in the weakly nonlinear regime when $Y(x,t) \approx 6$ close to the location of the maximum EPW amplitude, the variations of $v_g$ would mostly entail the defocussing of the wave packet, in contrast with the strongly nonlinear regime illustrated in Fig.~\ref{f1}. Finally, wherever $Y\agt 6$, $\nu$ is strongly reduced compared to its linear value while, as discussed above, $\partial_x v_g \leq 0$ in the rear side of the wave packet and, due to symmetric detrapping, $\partial_x v_g \approx 0$ in the front side. Hence, wherever $Y \agt 6$, the nonlinear and non local variations of $v_g$ are one main cause of dissipation, clearly more effective for longer and more intense wave packets. 

We therefore conclude that Landau damping, i.e.~the decay of the EPW amplitude along the characteristics, and the decrease of the EPW group velocity along the direction of propagation of the wave packet, are two distinct manifestations of collisionless dissipation, which are effective on distinct space domains. Moreover, the variations of $v_g$ may locally be anti-dissipative, which is most sensitive in the weakly nonlinear regime when $Y \approx 6$ close to the location of the EPW maximum amplitude.

\section{Multi-dimensional wave propagation}
  \label{III}
\subsection{Nonlinear envelope equation}
In this Paragraph, we use the same procedure as in the previous Section in order to generalize Eq.~(\ref{6})~to a multi-dimensional geometry. To do so, we just need to notice that, in Eq.~(\ref{6}), the term correcting Whitham's theory is proportional to the trapping rate, and, therefore, to the rate of variation of the EPW amplitude, calculated in the frame moving at the phase velocity with respect to the laboratory frame, and \textit{as seen by the electrons}. When all electrons has the same transverse velocity, $\vec{v}_\bot$, this term is nothing but $[\partial_\omega \chi_r^{\text{eff}}-\partial_\omega \chi_r][\partial_t E_p+v_\phi \partial_{x_{\sslash}} E_p +\vec{v}_\bot . \vec{\nabla}_\bot E_p]$, where, for paraxial wave propagation, $x_\sslash$ is along the local direction of the wave vector, and  $\vec{\nabla}_\bot$ is the gradient in the direction transverse to $\vec{k}$. The correcting term is easily generalized to allow for a warm distribution of transverse velocities so that, in more than one dimension, Eq.~({\ref{9})~becomes
\begin{equation}
\label{11}
\partial_t E_p+\left[v_\phi-\frac{2}{k \int f(\vec{v}_\bot) \partial_\omega \chi_r^{\text{eff}} d\vec{v}_\bot}Ê\right ]\partial_{x_\sslash}E_p 
+\frac{\left[\int \partial_\omega \chi_r^{\text{eff}} f(\vec{v}_\bot) \vec{v}_\bot d\vec{v}_\bot\right].\vec{\nabla}_\bot E_p }{\int f(\vec{v}_\bot) \partial_\omega \chi_r^{\text{eff}} d\vec{v}_\bot}+\nu E_p=0, 
\end{equation}
%\begin{equation}
%\label{11}
%\partial_t E_p+\left[v_\phi-\frac{2}{k \int f(\vec{v}_\bot) \partial_\omega \chi_r^{\text{eff}} d\vec{v}_\bot}Ê\right ]\partial_{x_\sslash}E_p 
%+\frac{\left[\int \partial_\omega \chi_r^{\text{eff}} f(\vec{v}_\bot) \vec{v}_\bot d\vec{v}_\bot\right].\vec{\nabla}_\bot E_p }{\int f(\vec{v}_\bot) \partial_\omega \chi_r^{\text{eff}} d\vec{v}_\bot}=\frac{E_d}{\int f(\vec{v}_\bot) \partial_\omega \chi_r^{\text{eff}} d\vec{v}_\bot}, 
%\end{equation}
where $f(\vec{v}_\bot)$ is the distribution of transverse velocities. Notice that, even when $f(\vec{v}_\bot)$ is isotropic, the third term in the left-hand side of Eq.~(\ref{11})~may be non zero because $\partial_\omega \chi_r^{\text{eff}}$ is a function of $\vec{v}_\bot$. Indeed, $\partial_\omega \chi_r^{\text{eff}}$~depends on the maximum wave amplitude, $E_{\max}$, \textit{experienced} by the electrons, and $E_{\max}$ is generally different, and usually larger, for the electrons exiting the wave packet than for those entering it. Hence, because the electron response to the wave is non local, the EPW group velocity may have a non zero transverse component, $\vec{v}_{g_\bot}$, which we now quantify in connection with the values assumed by the collisionless damping rate, $\nu$. In Ref.~\cite{benisti09}, we showed that electrons such that $Y_{\text{3D}}\equiv \int_{-\infty}^{t} \omega_B(x_\sslash-v_\phi t',\vec{x}_\bot-\vec{v}_\bot t',t') dt'Ê\agt 6$, i.e. those which have completed about one bounce period, no longer contribute to $\nu$, while all the others induce a nearly linear Landau damping of the EPW. This leads to the following estimate for $\nu$,
\begin{equation}
\nu=\nu_{\text{lin}} \frac{ \int f(\vec{v}_\bot) \partial_\omega \chi_r^{\text{eff}}\mathcal{H}\left[Y_{\text{3D}}\left(\vec{v}_\bot \right) \right] d\vec{v}_\bot}{ \int f(\vec{v}_\bot) \partial_\omega \chi_r^{\text{eff}}d\vec{v}_\bot},
\end{equation}
where $\nu_{\text{lin}}$ in the linear Landau damping rate and $\mathcal{H}(Y_{\text{3D}})$ is a Heaviside-like function, $\mathcal{H}(Y_{\text{3D}})\approx 0$ if $Y_{\text{3D}} \agt 6$ and  $\mathcal{H}(Y_{\text{3D}})\approx 1$ if $Y_{\text{3D}} \alt 6$. Similarly, in Ref.~\cite{benisti09} we showed that $\partial_\omega \chi_r^{\text{eff}}$ significantly departed from its linear value, $\partial_\omega \chi_r^{\text{lin}}$, only once $Y_{\text{3D}} \agt 6$, and that $\partial_\omega \chi_r^{\text{eff}}(Y_{\text{3D}}\agt 6) \gg \partial_\omega \chi_r^{\text{lin}}$. Moreover, when $Y_{\text{3D}} \agt 6$,  $\partial_\omega \chi_r^{\text{eff}}$ is essentially a decreasing function of $E_{\max}(x_\sslash,\vec{x}_\bot,t) \equiv \max_{t'<t}[E_p(x_\sslash-v_\phi t',\vec{x}_\bot-\vec{v}_\bot t',t') ]$. Using these results, we can now discuss the amplitude and orientation of $\vec{v}_{g_\bot}$.

In regions  where the wave amplitude has been so weak that $Y_{\text{3D}}<6$ for all electrons, $\nu \approx \nu_{\text{lin}}$ and $\partial_\omega \chi_r^{\text{eff}}=\partial_\omega \chi_r^{\text{lin}}$ independently of $\vec{v}_\bot$, so that $\vec{v}_{g_\bot}=\vec{0}$ if $f(\vec{v}_\bot)$ is isotropic. 

When the maximum wave amplitude has reached a large enough value, there may be space locations where $Y_{\text{3D}} > 6$ for the electrons exiting the plasma wave packet while $Y_{\text{3D}}<6$ for those entering it. There, for any $\vec{v}_\bot$ such that $\vec{v}_\bot.\vec{\nabla}E_p <0$, $\partial_\omega \chi_r^{\text{eff}}(\vec{v}_\bot) \gg \partial_\omega \chi_r^{\text{eff}}(-\vec{v}_\bot)$,  and $\vec{v}_{g_\bot} \propto \int \partial_\omega \chi_r^{\text{eff}} f(\vec{v}_\bot) \vec{v}_\bot d\vec{v}_\bot$ is clearly oriented towards the outside of the wave packet, which would tend to make it self-defocus. Moreover, in such space locations, about half of the electrons (those exiting the wave packet) do not contribute to the Landau damping rate and, therefore, $\nu \approx \nu_{\text{lin}}/2$. Note that, as the wave grows, the region where $Y_{\text{3D}}< 6$ for entering electrons is located further away from the center of the wave packet where the EPW amplitude is largest, so that only the edge of the plasma pulse is affected by the defocussing induced by $v_{g_\bot}$, which, therefore, becomes less effective. 

In space domains where $Y_{\text{3D}}>6$ for all electrons it is clear that $\nu \approx 0$. Moreover, since $E_{\max}$ is larger for electrons exiting the wave packet than for those entering it, and since, when $Y_{\text{3D}} \agt 6$, $\partial_\omega \chi_r^{\text{eff}}$ decreases with $E_{\max}$, $\partial_\omega \chi_r^{\text{eff}}(\vec{v}_\bot) < \partial_\omega \chi_r^{\text{eff}}(-\vec{v}_\bot)$ if $\vec{v}_\bot.\vec{\nabla}E_p <0$. Hence, in such space regions, $\vec{v}_{g_\bot}$ is oriented towards the inside of the wave packet and would tend to make it self-focus. However, since $\partial_\omega \chi_r^{\text{eff}}$ varies slowly with $E_{\max}$,  the magnitude of $\vec{v}_{g_\bot}$ is usually weaker when it induces a focussing effect than when it tends to make the wave packet self-defocus.

Let us now stress that $\vec{v}_{g_\bot}$ is easily connected to collisionless dissipation since it is non zero just because, due to symmetric detrapping, $\partial_\omega \chi_r^{\text{eff}}$ is a non local function of the EPW amplitude. As for collisionless dissipation, in the regime when $\omega_B \gg \gamma$, it is also only due to symmetric detrapping, as discussed in the previous Section. Actually, from Eq.~({\ref{11})~one easily finds
\begin{equation}
\label{11a}
\frac{d}{dt}Ê\int E_p^2 d^3r=-2 \int \nu  E_p^2 d^3r+\int \partial_{x_\sslash} v_{g_\sslash} E_p^2 d^3r + \int \vec{\nabla}_\bot. \vec{v}_{g_\bot} E_p^2 d^3r,
\end{equation}
where $v_{g_\sslash}$ is the component of the group velocity along the $x_\sslash$ direction [given by the second term in the lest-hand side of Eq.~(\ref{9})]. The effect of $\partial_{x_\sslash} v_{g_\sslash}$ on the time variation of the electrostatic energy is exactly the same as in the one-dimensional situation. For each $(\vec{x}_\bot,t)$, wherever $Y_{\text{3D}}(x_\sslash,\vec{x}_\bot,t) \alt 6$, $\partial_{x_\sslash} v_{g_\sslash} \approx 0$, wherever $Y_{\text{3D}}(x_\sslash,\vec{x}_\bot,t)\agt 6$, $\partial_{x_\sslash} v_{g_\sslash} < 0$ in the rear side of the wave packet and $\partial_{x_\sslash} v_{g_\sslash} \approx 0$ in the front side, and in the small domain where $Y_{\text{3D}}(x_\sslash,\vec{x}_\bot,t) \approx 6$, $\partial_{x_\sslash} v_{g_\sslash} > 0$. Hence, the effect of $\partial_{x_\sslash} v_{g_\sslash}$ is mostly dissipative in the strongly nonlinear regime, for long and intense pulses, and anti-dissipative in the weakly nonlinear regime when $Y_{\text{3D}}\approx 6$ close to the maximum wave amplitude. It is noteworthy that the same conclusion holds for each $ \vec{\nabla}_\bot. \vec{v}_{g_\bot}(\vec{v}_\bot)$, where $\vec{v}_{g_\bot}(\vec{v}_\bot)$ is the contribution to $\vec{v}_{g_\bot}$ of those electrons which all have the same transverse velocity, $\vec{v}_\bot$. Indeed, for each $(x_\sslash,t)$,  wherever $Y_{\text{3D}}(x_\sslash,\vec{x}_\bot,t) \alt 6$, $\vec{\nabla}_\bot. \vec{v}_{g_\bot}(\vec{v}_\bot) \approx 0$, wherever $Y_{\text{3D}}(x_\sslash,\vec{x}_\bot,t)\agt 6$, $\vec{\nabla}_\bot. \vec{v}_{g_\bot}(\vec{v}_\bot) < 0$ if $E_p<E_{\max}$ and $\vec{\nabla}_\bot. \vec{v}_{g_\bot}(\vec{v}_\bot)\approx 0$ if $E_p=E_{\max}$, and in the small domain where $Y_{\text{3D}}(x_\sslash,\vec{x}_\bot,t) \approx 6$, $\vec{\nabla}_\bot. \vec{v}_{g_\bot}(\vec{v}_\bot) > 0$. Hence, in the strongly nonlinear regime the effect of $\vec{\nabla}_\bot. \vec{v}_{g_\bot}$ is mostly dissipative, while it is mostly anti-dissipative in the weakly nonlinear regime. However, whatever the nonlinear regime, weak or strong, the anti-dissipative effect of $\vec{\nabla}_\bot. \vec{v}_{g_\bot}$ always shows up because there is always a region where $Y_{\text{3D}} > 6$ for exiting electrons while $Y_{\text{3D}}<6$ for entering electrons. There, $\vec{v}_{g_\bot}(\vec{v}_\bot)$ is much larger in magnitude than $\vec{v}_{g_\bot}(-\vec{v}_\bot)$ if $\vec{v}_\bot.\vec{\nabla}_\bot E_p <0$, so that the effect of $\vec{v}_{g_\bot}$ is defocussing. In the weakly nonlinear regime, defocussing is effective near the center of the wave packet, where the EPW amplitude is largest. However, as the wave grows, it is clear that the defocussing region moves towards the edge of the wave packet and, therefore, becomes less effective. %Moreover, in the strongly nonlinear regime, the orientation of $\vec{v}_{g_\bot}$ near the center of the wave packet would tend to make it self-focus, as already discussed above. 

It is noteworthy that an analysis similar to ours, on the net energy taken by the electrons from the wave due to trapping, was already made in Section IV of Ref.~\cite{banks}~by Banks \ea, who insisted on the fact that collisionless dissipation was in competition with the EPW focussing due to wave front bowing. Namely, they wrote that ``the focussing effect of curved wave fronts is limited by (\dots) the loss of energy field to resonant electrons that transit the wave\dots''. However, in Ref.~\cite{banks}, collisionless dissipation was identified with Landau damping while, here, we show that in the weakly nonlinear regime, collisionless dissipation may entail a defocussing effect for the EPW, independently of Landau damping. Note that, in Ref.~\cite{banks}, was also discussed the \textit{linear} defocussing of the EPW due a diffraction-like effect which we do not address here. This issue is postponed to a forthcoming paper, where we derive an envelope equation for the EPW at second order in its space variations. 

\subsection{Results from simulations of stimulated Raman scattering using the envelope code BRAMA}

In order to quantify the defocussing effect due to collisionless dissipation and to compare it with the focussing induced by wave front bowing, we now present results obtained with the envelope code BRAMA for an EPW driven by the optical mixing of two counterpropagating lasers. The code BRAMA, described in detail in Ref.~\cite{brama}, was recently generalized to allow for a three-dimensional space variation of the wave amplitudes, and, as regards the electromagnetic fields, it just solves the equations of Ref.~\cite{brama}~with an additional term accounting for diffraction. As for the electrostatic wave, its envelope equation is just a mere generalization of Eq.~(\ref{9})~obtained by allowing for the coupling to the electromagnetic waves.  Then, the equation for the EPW amplitude is
\begin{equation}
\label{12bis}
\partial_t E_p+\vec{v}_g \vec{\nabla}E_p+\nu E_p=\frac{E_d}{ \partial_\omega \chi_{\text{3D}}^{\text{eff}}},
\end{equation}
where $E_d$ accounts for the effect of the laser drive and where we denoted
\begin{equation}
\label{15}
\partial_\omega \chi_{\text{3D}}^{\text{eff}} \equiv \int f(\vec{v}_\bot) \partial_\omega \chi_r^{\text{eff}} d\vec{v}_\bot.
\end{equation}
We restrict here to paraxial wave propagation along the $x$-direction, and approximate $\vec{\nabla}_\bot E_p$~by $\partial_y E_p \hat{y}+ \partial_z E_p \hat{z}$. Then, $v_g.\vec{\nabla} E_p \equiv v_{g_x}\partial_x E_p +v_{g_y}\partial_y E_p +v_{g_z}Ê\partial_z E_p$ with
\begin{eqnarray}
\label{13}
v_{g_x}Ê&\approx &v_\phi-\frac{2}{k \partial_\omega \chi_{\text{3D}}^{\text{eff}}Ê}\\
\label{14}
v_{g_{y,z}}& = &\left[v_\phi-\frac{2}  {k \partial_\omega \chi_{\text{3D}}^{\text{eff}} } \right ]\frac{k_{y,z}}{k} +\frac{\int \partial_\omega \chi_r^{\text{eff}} f(\vec{v}_\bot) v_{y,z} d\vec{v}_\bot}{  \partial_\omega \chi_{\text{3D}}^{\text{eff}}  },
\end{eqnarray}
where the transverse components of the wave vector are derived using the consistency relation (see for example Ref.~\cite{whitham})
\begin{equation}
\label{16}
\partial_t \vec{k}=-\vec{\nabla} \omega = -\frac{d\omega}{dE_p} \vec{\nabla} E_p.
\end{equation}
The nonlinear dependence of $\omega$ with respect to $E_p$ may be found in Refs.~\cite{benisti08,lindberg}, showing that ${d\omega}/{dE_p}$ is usually negative. Hence, the first term in the right-hand side of Eq.~(\ref{14}) is directed along the transverse gradient of the EPW amplitude, i.e., towards the inside the wave packet. Therefore, the nonlinear frequency shift, which entails the wave front bowing (see Ref.~\cite{yin08}),  tends to let the plasma wave packet self-focus. 

\begin{figure}[!h]
\centerline{\includegraphics[width=11cm]{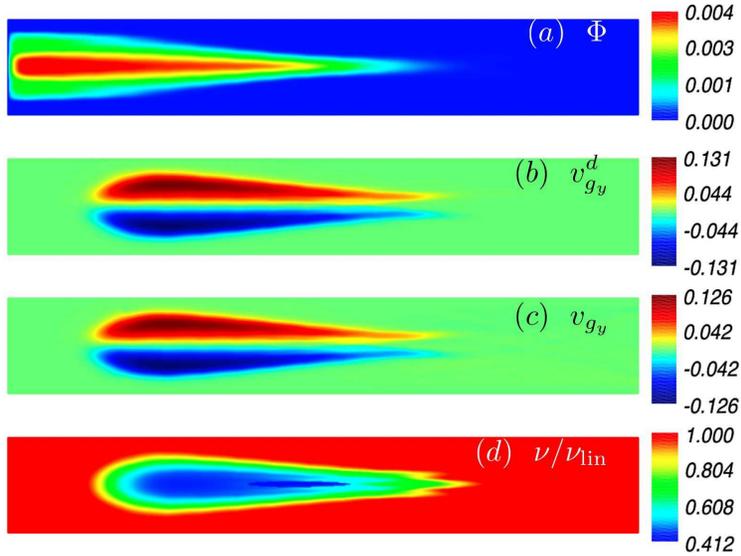}}
\caption{\label{f3}(Color online). Results obtained with BRAMA, at time $t=1$ps, showing the profiles of: panel (a), $\Phi \equiv eE_p/kT_e$ ; panel (b), the transverse component of the group velocity, $v_{g_y}^d$, induced by collisionless dissipation, and normalized to the modulus of the group velocity, $v_g$ ; panel (c), the total transverse component, $v_{g_y}$, of the EPW group velocity, normalized to $v_g$ ; panel (d), the collisionless damping rate normalized to its linear value. At time $t=1$ps, when the Landau damping rate is still significant compared to its linear value, $v_{g_y}\approx v_{g_y}^d$, and the sign of $v_{g_y}$ leads to an effective defocussing effect.}
\end{figure}
\begin{figure}[!h]
\centerline{\includegraphics[width=11cm]{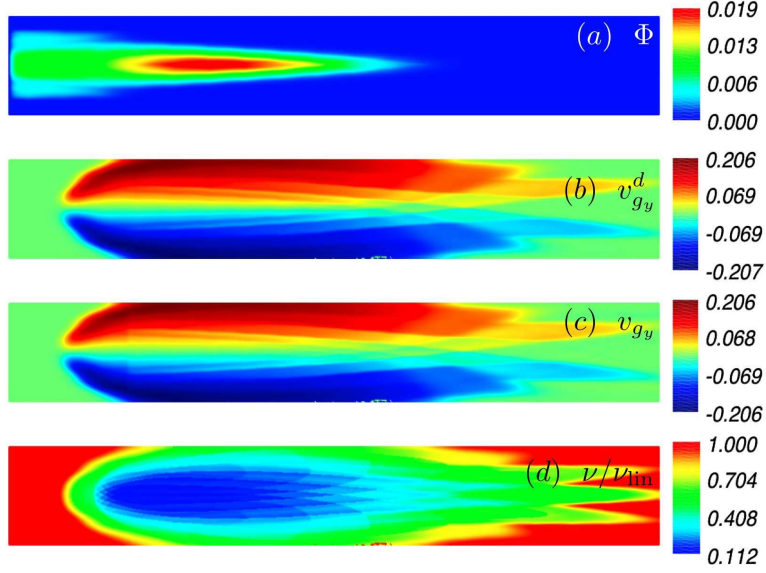}}
\caption{\label{f4}(Color online). Same as Fig.~\ref{f3} but at time $t=1.5$ps. As in Fig.~\ref{f3}, the defocussing effect due to collisionless dissipation is dominant. Note that, in spite of this, the transverse profile of the plasma wave packet has shrunk compared to the previous Figure because of the inhomogeneity of the SRS growth rate.}
\end{figure}
\begin{figure}[!h]
\centerline{\includegraphics[width=11cm]{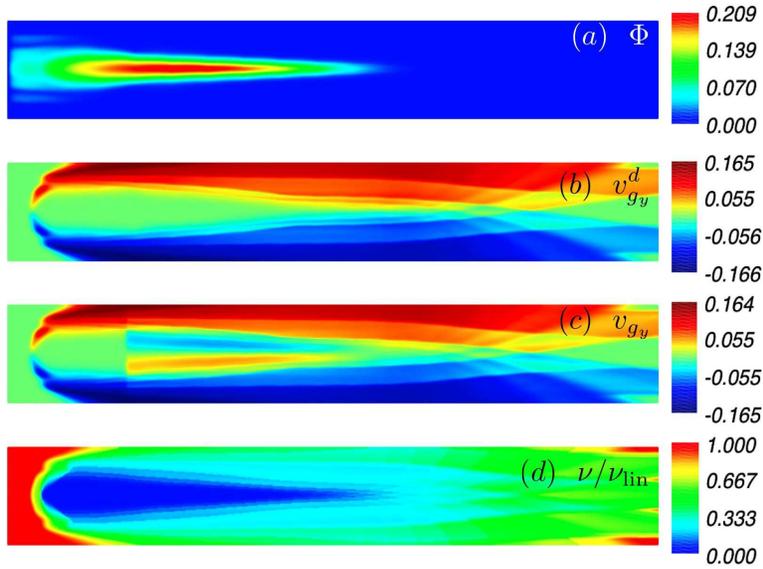}}
\caption{\label{f5}(Color online). Same as Fig.~\ref{f3} but at time $t=2$ps, when Landau damping has nearly vanished. At this time, the focussing effect due to wave front bowing prevails near the axis and at the front side of the wave packet.}
\end{figure}
\begin{figure}[!h]
\centerline{\includegraphics[width=11cm]{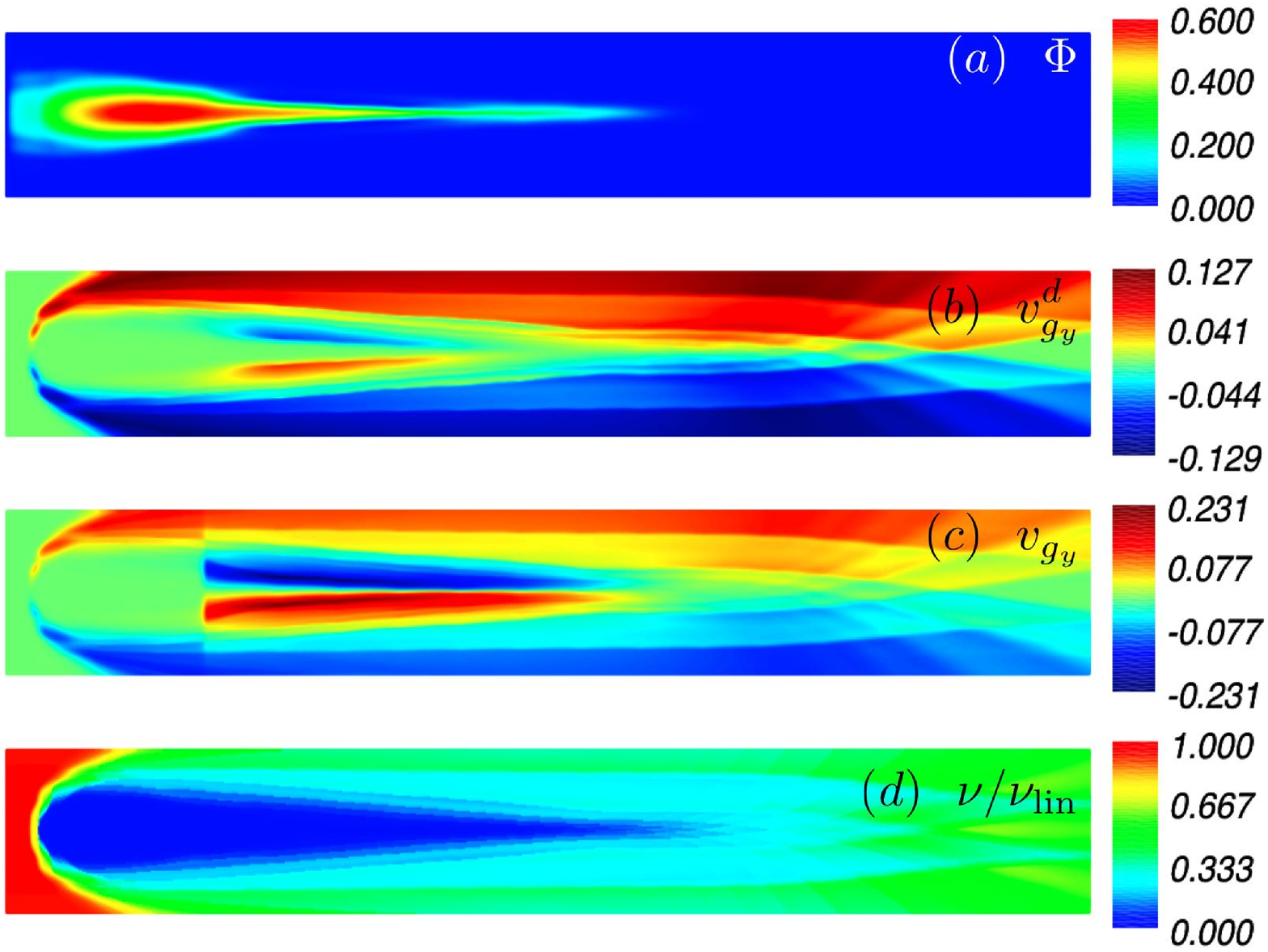}}
\caption{\label{f6}(Color online). Same as Fig.~\ref{f3} but at time $t=2.3$ps, when the component $v_{g_y}^d$ due to dissipation has changed sign close to the axis and at the front side of the wave packet, which keeps on focussing more rapidly. Note the change in shape of the EPW profile between Figs.~\ref{f5}~and~\ref{f6}, resulting from the focussing of the front side.}
\end{figure}
In this paper, we restrict to an envelope equation at first order in the space derivatives of the EPW amplitude. The inclusion of a diffraction-like effect, as that discussed by Banks \ea~in Ref.~\cite{banks}, is postponed to a forthcoming paper where we will allow for \textit{nonlinear}~terms proportional to the second order space derivatives of the EPW amplitude. These terms are however not essential for a plasma wave driven by stimulated Raman scattering (SRS), considered in this Subsection, because an electrostatic wave is much less subjected to diffraction than electromagnetic ones since it has a much smaller group velocity. Discussing this in detail is however way beyond the scope of this paper, and will be addressed elsewhere.  

Figs.~\ref{f3}~to~\ref{f6} show results from a two-dimensional simulation of the optical mixing between a laser with intensity $I_l=6\times10^{15}$W/cm$^2$ and a counterpropagating  ``seed'' with intensity $I_s=6\times10^{5}$W/cm$^2$, inside a plasma whose electron temperature is $T_e=0.7$keV and electron density is $n_e=4\times10^{-2}n_c$, where $n_c$ is the critical density. The simulation domain is $140\mu$m long and $14\mu$m wide along the $y$-direction. The laser wavelength is $0.527\mu$m and, at best focus (located in the middle of the simulation box), its envelope is a Gaussian with intensity varying as $\exp(-y^2/w^2)$, with a waist $w= 1\mu$m. The seed wavelength is $\lambda_s= 0.689\mu$m and its phase is uniform at the entrance of the simulation box, which is discretized using 1000 points in the $x$-direction and 128 points in the $y$-direction. The distribution of the transverse electron velocity is assumed to be a Maxwellian, which we discretize over 25 evenly spaced values from $-4 v_{th}$ to $4 v_{th}$, where $v_{th}$ is the thermal velocity.   

As shown in Figs.~\ref{f3}~and~\ref{f4}, in the weakly nonlinear regime, when the Landau damping rate is still significant compared to its linear value, the transverse component of the group velocity induced by collisionless dissipation, i.e. the last term in the left-hand side of Eq.~(\ref{14})~which we denote by $v_{g_y}^d$, entails a  defocussing effect, in agreement with the discussion of the previous Paragraph. Moreover, it is easily checked that $v_{g_y}^d$ is about ten times larger than the linear speed of defocussing due to diffraction-like effects, which, from the formulas of Ref.~\cite{banks}, would be $v_{g_y}^{\text{lin}} \approx 8\times10^{-3} v_g$ at time $t=1$ps. We also note that defocussing is effective near the center of the wave packet, although, exactly at the center, $v_{g_y}^d\approx0$ since $E_{\max}$ is the same for entering and exiting electrons. In Figs.~\ref{f3}~and~\ref{f4}, the total transverse group velocity, $v_{g_y}$ has the same orientation as $v_{g_y}^d$, showing that the defocussing effect induced by $v_{g_y}^d$ prevails over the focussing effect due to wave front bowing. In spite of this, the transverse extent of the plasma wave packet decreases because the SRS growth rate is larger close to the axis than near the edge of the simulation box. As the wave keeps on growing, the regime of wave-particle interaction becomes more nonlinear and the Landau damping rate keeps on decreasing where the EPW amplitude is significant. Then, as discussed in the previous Paragraph, the defocussing region moves towards the edge of the plasma pulse so that, eventually, the focussing effect due to wave front bowing dominates near the center and at the front side of the plasma wave packet, as shown in Fig.~\ref{f5}. This clearly affects the profile of the EPW, as may be seen by comparing the results of Fig.~\ref{f5}~to those of Fig.~\ref{f6}.  Finally, in the strongly nonlinear regime, when $\nu \approx 0$ where the EPW amplitude is significant, $v_{g_y}^d$ changes sign close to the axis and on the front side of the wave packet, which therefore keeps on strongly focussing.

\section{Conclusion}
\label{conc}
In this paper, we discussed in detail the distinction that may exist between Landau damping and collisionless dissipation. Three regimes of wave-particle interaction actually need to be considered. When the electron response to an EPW is well approximated by linear theory, the wave is Landau damped, its amplitude decreases along the characteristics, while its group velocity remains constant and uniform. In the linear regime, dissipation is only due to Landau damping. In the nonlinear regime and in one dimension, the EPW group velocity, $v_g$, remains nearly  at its linear value as long as $Y\equiv\int_{-\infty}^t \omega_B(x-v_\phi t',t') dt'  \alt 6$, and quickly increases towards the phase velocity where $Y \approx 6$, i.e.~where the trapped electrons have completed about on bounce period, which entails a local increase in the size of the wave packet. For larger values of $Y$, $v_g$ decreases with $x$ (i.e.~along the direction of propagation of the wave packet), until the maximum wave amplitude is reached, and then remains constant. This clearly induces the shrinking of the wave packet in that region located between $x=x_6$, where $Y \approx 6$, and $x=x_{\max}$ where the EPW amplitude reaches its maximum value. In the weakly nonlinear regime, when $Y\approx6$ close the $x=x_{\max}$, $v_g$ increases where the EPW amplitude is largest and then remains constant for $x>x_{\max}$. Hence, the nonlinear variations of $v_g$ lead to an increase in the size of the wave packet, which is an anti-dissipative effect. As for the collisionless damping rate, it remains nearly at its linear value for $x<x_{\max}$ so that, again, damping is the only dissipative mechanism. By contrast, in the strongly nonlinear regime, when $Y \approx 6$ in the rear side of the wave packet where the wave amplitude is very small compared to its maximum value, only \textit{there} is the damping rate significant compared to its linear value, so that damping is much less effective than in the linear or weakly nonlinear regimes. As for the nonlinear variations of $v_g$, they lead to the shrinking of the wave packet in the region $x_6<x<x_{\max}$, where the EPW amplitude is significant, which automatically reduces the electrostatic energy $W \equiv \int_{-\infty}^{+\infty}ÊE_p^2dx$. This a dissipative mechanism different from Landau damping, and more effective for longer and more intense wave packets. Physically, it is directly due to trapping, which is dissipative, since the energy gained by the electrons while being trapped is not given back to the wave when they are \textit{symmetrically}~detrapped. In Ref. \cite{vgroup}, we wrote that $v_g \neq d\omega/dk$ due to a term in the envelope equation that may be viewed as the nonlinear counterpart of the Landau damping rate, although the result  $v_g \neq d\omega/dk$ was only valid once Landau damping has nearly vanished. Hopefully, in this paper, we clarified this statement by showing that the nonlinear variations of the EPW group velocity, as derived in Ref.~\cite{vgroup}, was one manifestation of collisionless dissipation  effective in the strongly nonlinear regime, i.e., after Landau damping has nearly vanished where the wave amplitude is significant. 

The other important point of this paper was to recognize how transparent was,  in the envelope equation for the EPW amplitude, the effect of trapping on collisionless dissipation. Once this is understood, the envelope equation derived when the wave amplitude is uniform is straightforwardly generalized to allow for three-dimensional effects. The conclusions are essentially the same as in one dimension. In the weakly nonlinear regime, dissipation is only due to Landau damping, while the increase of the group velocity compared to its linear value would induce an anti-dissipative effect by increasing the space extent of the wave packet. In the strongly nonlinear regime, damping is strongly reduced, and the nonlinear variations of the group velocity entails a shrinking of the wave packet and, therefore, of the electrostatic energy. However, we came to the quite unexpected result that dissipation induced a transverse component to the EPW group velocity, which we precisely quantified. In the weakly nonlinear regime, this transverse component is directed towards outside the wave packet, leading to a defocussing effect that might overcome the focussing due to wave front bowing, as illustrated by results from two-dimensional simulations of optical mixing using the envelope code BRAMA. As the EPW amplitude keeps increasing, and the wave-particle interaction becomes more non linear, the transverse component of the group velocity due to dissipation eventually entails a focussing effect that reinforces that due to bowing. 

At this stage, we would like to point out that, as clearly shown in Refs.~\cite{brunner,vlasovia}, symmetric detrapping leads to a positive slope in the electron distribution function, that may entail the unstable growth of electrostatic modes. This may be viewed as an indirect consequence of collisionless dissipation, which will however not be discussed in this paper. This is left for future work.  


\begin{thebibliography}{10}
\bibitem{note}
As in Ref.~\cite{vgroup}, we are using here an approximate version of Whitham's variational approach, neglecting the time and space derivatives of $k$ and $\omega$, whose impact on the group velocity, for example, is beyond the scope of this paper devoted to the effect of collisionless dissipation. The numerical results of Refs.~\cite{benisti09,vgroup,brama} showed that our approximation was quite accurate, both as regards the free propagation of a plasma wave packet and the application to stimulated Raman scattering, at least for the physics parameters chosen in the cited papers, which are relevant to inertial confinement fusion.%We may therefore miss  effects such as group velocity splitting, or unstable wave propagation, all clearly explained in Chapter 14 of Whitham's book, Ref.~\cite{whitham}, and in Ref.~\cite{dodin}. When collisionless dissipation is large, as for the parameters of Figs.~\ref{f3}-\ref{f6}, unstable wave propagation seems to be the dominant effect (see Ref. \cite{note2}), but addressing this is beyond the scope of this article devoted to collisionless dissipation (see the related discussion at the end of the conclusion).
\end{thebibliography}
\end {document}